\documentclass[letterpaper]{article}
\usepackage{flairs}%aaai
\usepackage{times}
\usepackage{helvet}
\usepackage{courier}

\usepackage{amsmath, amsfonts, amssymb, amsthm, natbib}
\usepackage{graphicx,psfrag,epsf}
\usepackage{latexsym}
\usepackage{calc}
\usepackage[ragged, symbol]{footmisc}
\usepackage{grffile}
\usepackage{booktabs}
\usepackage{algorithmic}
\usepackage{algorithm}
\usepackage{caption}
\usepackage{comment}
\usepackage{color,url}
\usepackage{enumitem}
\usepackage{bbm}

\usepackage{graphicx,psfrag,epsf}
\usepackage{bm,multirow}
\usepackage{natbib}
\usepackage{url} % not crucial - just used below for the URL 

\usepackage[utf8]{inputenc} % allow utf-8 input
\usepackage[T1]{fontenc}    % use 8-bit T1 fonts
\usepackage{hyperref}       % hyperlinks
\usepackage{url}            % simple URL typesetting
\usepackage{booktabs}       % professional-quality tables
\usepackage{amsfonts}       % blackboard math symbols
\usepackage{nicefrac}       % compact symbols for 1/2, etc.
\usepackage{microtype}      % microtypography
 
\usepackage{float}
\usepackage{txfonts}

\newcommand{\Real}{\mathbb{R}}

\newcommand{\Tra}{^{\sf T}} % Transpose
\def\vec{\mathop{\rm vec}\nolimits}

\newcommand{\tr}{\operatorname{tr}} % Trace
\newcommand{\V}[1]{{\bm{\mathbf{\MakeLowercase{#1}}}}} % vector
\newcommand{\M}[1]{{\bm{\mathbf{\MakeUppercase{#1}}}}} % matrix

\begin{document}
\title{A Regularized Vector Autoregressive Hidden Semi-Markov model, with Application to Multivariate Financial Data}
\author{Zekun Xu \\
  North Carolina State University \\
  Raleigh, NC USA\\
  \texttt{\small zxu13@ncsu.edu} \\\And
  Ye Liu \\
  North Carolina State University \\
  Raleigh, NC USA\\
  \texttt{\small yliu87@ncsu.edu}}
\maketitle
\begin{abstract}
\begin{quote}

Hidden Markov model (HMM) has been a popular choice for financial time series modeling due to its advantage in capturing dynamic regimes. However, HMM's implicit assumption that the state duration follows a geometric distribution is too strong to hold in practice. 
In this work, we propose a regularized vector autoregressive hidden semi-Markov model to analyze multivariate financial time series.
One challenge in such a model setting is that the number of parameters is too large to be reliably estimated unless the time series is extremely long. To address this issue,
an augmented EM algorithm is developed for parameter estimation by using regularized estimators for the state-dependent covariance matrices and autoregression matrices in the M-step. 
The performance of the proposed model is evaluated in a simulation experiment, and demonstrated with the New York Stock Exchange financial portfolio data.
\end{quote}
\end{abstract}

\section{Introduction}

In finance and economics, it is often assumed that the financial returns follow a white noise process. However, empirical evidence suggests that this assumption may be too strong to hold in practice. \cite{ding1993long} found that there is substantial correlation between absolute returns. \cite{andersen2001distribution} indicated that realized volatilities and correlations show strong temporal dependence and appear to be well described by long-memory processes. Moreover, \cite{fan2017elements} commented that the squared and the absolute returns of S\&P 500 index exhibits significant serial correlations. Therefore, it is reasonable to model the financial return series using an autoregressive process.

The drawback of an autoregressive process is that it alone cannot model the volatility clustering and heavy-tailed distribution in the financial return series. This is because such financial return series often have more than one latent data generating mechanisms. For example, the performance of a financial portfolio in a stable economy is expected to follow a different autoregressive process from that in a volatile economy. \cite{ryden1998stylized} showed that a hidden Markov model (HMM) can reproduce most of the stylized facts for daily return series \cite{granger1995some}.

HMM is a bivariate discrete time stochastic process $\{S_t,Y_t\}_{t\geq 0}$ such that

\begin{enumerate}[topsep=0pt,itemsep=-1ex,partopsep=1ex,parsep=1ex,label=(A\arabic*)]
\item $\{S_t\}$ is a Markov chain, i.e. $P(S_t|S_{t-1},...,S_1)=P(S_t|S_{t-1})$.
\item $\{Y_t\}$ is a sequence of conditional independent random variables given $\{S_t\}$.
\end{enumerate} 

In a Gaussian HMM, the marginal distributions for observed series are essentially modeled as a mixture of Gaussian distributions such that volatility clustering and heavy-tailedness are automatically incorporated in the model framework. Further, the transition between the latent states are directly modeled in HMM so as to account for the temporal dependence in the series. 

However, assumptions (A1) and (A2) may both be too strong to hold in practice. Assumption (A1) indicates that the current latent state depends only on the most recent latent state in the past; beyond that, it is memoryless. \cite{ryden1998stylized} illustrated that the stylized fact of the very slowly decaying autocorrelation for absolute (or squared) returns cannot be described by a HMM. \cite{bulla2006stylized} proposed the use of hidden semi-Markov model (HSMM) to overcome the lack of flexibility of HMM to model the temporal higher order dependence in financial returns. In HSMM, the latent state durations are explicitly modeled rather than assuming them to be geometric as in HMM. This has the practical advantage since it is typical that the longer time the economy spends in one of the latent states the more likely it will switch to another latent state. In the meantime, assumption (A2) can be dropped in the class of Markov-switching models proposed by \cite{hamilton1989new} where $\{Y_t\}$ is allowed to follow state-dependent Gaussian vector autoregressive processes, also known as vector autoregressive hidden Markov models (VAR-HMM). \cite{yang2000some} pointed out another interesting feature that VAR-HMM can occasionally behave in a nonstationary manner although being stationary and mean reverting in the long run. 

For general applicability, we are going to adopt the most flexible framework of a $p^{th}$ order vector autoregressive hidden semi-Markov model [VAR(p)-HSMM] to analyze multivariate financial time series. Note that both VAR and HMM are special cases in the VAR(p)-HSMM framework. Our goal is to make inference on the parameters that determine the data generating mechanism, as well as evaluate the prediction performance. A potential problem of VAR(p)-HSMM is the large number of parameters to be estimated when the dimension of $Y_t$ is high. A multivariate M-state VAR(p)-HSMM series of dimension $d$ has $\frac{Md(d+1)}{2}$ parameters in the state-dependent covariance matrices and $Mpd^2$ parameters in the autoregression matrices. Unless the time series is extremely long, we are not able to reliably estimate the covariance and autoregression matrices even when the dimension $d$ is moderate. Therefore, regularizations are needed to stabilize the parameter estimation. \cite{stadler2013penalized}, \cite{fiecas2017shrinkage}, and \cite{monbet2017sparse} proposed different versions of a penalized log-likelihood procedure with regularization on the state-dependent inverse covariance matrices in a Gaussian HMM to form a more stable regularized estimator. So far, there is no literature that elaborates on the regularized estimation for VAR(p)-HSMM framework. Neither has the regularized VAR(p)-HSMM framework been used to model multivariate financial returns yet.

Thus, our contribution is to provide a detailed parameter estimation procedure for a regularized VAR(p)-HSMM. The model framework of VAR(p)-HSMM is provided in Section 2, where we integrated the LASSO regularization by \cite{tibshirani1996regression} on autoregression matrices, and shrinkage regularization by \cite{ledoit2004well} on covariance matrices into the EM algorithm. Section 3 presents simulation studies on finite samples to evaluate the performance of the proposed regularized estimators in different scenarios. Section 4 provides an empirical analysis on the NYSE financial portfolio of 50 stocks using the regularized VAR(p)-HSMM. Section 5 gives a brief discussion. All the analyses utilize the R package "rarhsmm", which has been developed for fitting regularized VAR(p)-HSMM.

\section{Methodology}\label{sec:methodology}

\subsection{Model framework}

Denote by $\V{Y}_t \in \Real^d$ for t=1,...,T to be the observed multivariate data at time t, where d is the dimension for each $\V{Y}_t$. Denote by $S_t\in \{1,...,M\}$ to be the latent state at time t, where M is the fixed finite number of states. Let $\V{\delta}=[\delta_1,...,\delta_M]$ be the prior probability of latent states. Further, we denote the latent state duration densities by $\V{r}=[r_1,...,r_M]$ such that
\begin{equation}
r_{i}(n)=P(\textrm{stay n times in latent state i}) \;\;\;\;\;\;\;\;\;\;n=1,2,...,D, \nonumber
\end{equation} 
where D is the fixed maximum state duration, i.e. any state duration greater than D will be censored at D. In addition, denote by $\M{Q}=\{q_{ij}\}$ for i=1,...,M and j=1,...,M the state transition matrix such that
\begin{equation}
q_{ij}=P(S_{t+1}=j|S_t=i) \;\;\;\;\;\;\;\;\;\;t=1,...,T-1, \nonumber
\end{equation} 
where $\sum_{j=1}^Mq_{ij}=1\;\;\;\forall\;\;i \in {1,...,M}$

Thus, the data generating mechanism for VAR(p)-HSMM, can be described as follows. First, an initial state, $S_1=i\;(i \in {1,...,M})$ is chosen according to the initial state distribution $\delta_i$. Second, a duration $n$ is chosen according to the latent state duration density $r_{i}(n)$. Third, observations $\V{Y}_1,...,\V{Y}_{n} \in \Real^d$ are chosen according to the state-dependent $p^{th}$ order Gaussian vector autoregressive process  
\begin{equation}
\V{Y}_t=\V{\mu}_i + \sum_{k=1}^p\M{A}_{ki}\V{Y}_{t-k}  + \V{\epsilon}_{ti} \textrm{  where  }\V{\epsilon}_{ti} \sim \mathcal{N}(\V{0},\M{\Sigma}_i),
\end{equation}
for i = 1,...,M and t = 1,...,n, where $\V{\mu}_i \in \Real^d$ and $\M{\Sigma}_i \in \Real^{d\times d}$ are the conditional mean and covariance matrix of $\V{Y}_t$ given $S_t,\V{Y}_{t-1},...,\V{Y}_{t-p}$; $\M{A}_{ki} \in \Real^{d\times d}$ is the $k^{th}$-order autoregression matrix conditioning on $S_t=i$. 

Fourth, the next state, $S_{n+1}=j$, is chosen according to the state transition probability $q_{ij}$, the $i,j^{th}$ element in the transition matrix $\M{Q}$. An implicit constraint is that there should be no transition back to the same state because we generate exactly n observations in latent state i in the previous steps, i.e. $S_{1:n}=i$. Then the data generating process repeats the previous steps until we end up with T observations.

Denote by $\V{\theta}=[\V{\delta},\V{r},\M{Q},\V{\mu},\M{\Sigma},\M{A}]$ the set of all parameters in VAR(p)-HSMM, where there are $M-1$ free parameters in $\V{\delta}$, $M(D-1)$ in $\V{r}$, $M(M-2)$ in $\M{Q}$, $Md$ in $\V{\mu}$, $\frac{Md(d+1)}{2}$ in $\M{\Sigma}$, and $Mpd^2$ in $\M{A}$.
 
Our VAR(p)-HSMM framework is a natural generalization of the VAR(p)-HMM framework \citep[][]{hamilton1989new,yang2000some,monbet2017sparse,francq2001stationarity}  by allowing for the explicit modeling of the state duration distributions. In particular, we set all the latent state duration densities to be discrete nonparametric distributions with arbitrary point mass assigned to the feasible duration values so as to allow for the most flexibility. 
%Melnyk et al.(2016) [10] used VAR-HSMM for anomaly detection in aviation systems, but so far no regularization methods have been discussed for VAR-HSMM.

\subsection{Regularization}

There are two motivations for us to apply regularization on the VAR(p)-HSMM framework. On the one hand, the daily financial time series is typically not long enough for us to reliably estimate all the parameters in the state-dependent covariance matrices in the VAR(p)-HSMM. Those covariance matrices may not be invertible especially when the dimension of $\V{y}_t$ is high. On the other hand, we assume that the state-dependent autoregression matrices to be sparse, i.e. many entries are nearly zero.
Although the white noise assumption is often used in financial return data, the empirical evidence indicates that
the IID assumption is too strong and too restrictive to be true in general \citep[][]{fan2017elements,franke2004statistics}. Thus, a regularized estimator for autoregression matrices can shrink the negligible correlations to zero while allow for the possibility that some correlations may be significant.
%\cite{tsay2005analysis} p36 https://books.google.com/books?id=OKUGARAXKMwC&pg=PA36&lpg=PA36&dq=financial+return+white+noise&source=bl&ots=tpaEIbvuxs&sig=FV47_p4Ui9JUsrf2aS8pw0aMBIk&hl=en&sa=X&ved=0ahUKEwjP9vzE2YXXAhWCSSYKHUdECjwQ6AEILDAB#v=onepage&q=financial%20return%20white%20noise&f=false
%jurgen p263   https://books.google.com/books?id=jtx5BgAAQBAJ&pg=PA264&lpg=PA264&dq=financial+return+white+noise&source=bl&ots=FyZBtOtGXS&sig=Qg1nIjtgsETE3vujHniOAsn0MTw&hl=en&sa=X&ved=0ahUKEwjP9vzE2YXXAhWCSSYKHUdECjwQ6AEIUzAJ#v=onepage&q=financial%20return%20white%20noise&f=false 
%fan: chapter 1 p18   http://orfe.princeton.edu/~jqfan/fan/FinEcon/chap1.pdf

The regularized estimator for state-dependent covariance matrices follows the work of \cite{ledoit2004well}, \cite{sancetta2008sample}, and \cite{fiecas2017shrinkage} such that each regularized estimator is a convex combination of the maximum likelihood estimator and a scaled identity matrix with the same trace,
$$\M{\Sigma}^r=\frac{1}{1+\lambda_\Sigma}\hat{\M{\Sigma}}^{mle}+\frac{\lambda_\Sigma}{1+\lambda_\Sigma}c\M{I} \;\;\;\;s.t\;\;\;\; \tr(\hat{\M{\Sigma}}^{mle})=\tr(c\M{I}),$$
where $\lambda_\Sigma \geq 0$ controls the strength of the regularization. Note that when $\lambda_\Sigma=0$, we have $\M{\Sigma}^r=\hat{\M{\Sigma}}^{mle}$.
This regularized estimator results in shrinkage on the covariance estimates and ensures the positive definiteness of the estimated covariance matrix when the sample covariance matrix is close to singularity. This holds even if $\lambda_\Sigma$ is very small so that we do not increase much bias when stabilizing the estimate. Besides, this regularization yields not only invertible but also well-conditioned covariance estimates. As $\lambda_\Sigma$ increases, the dispersion between the smallest and the largest eigenvalues for the estimated covariance matrix shrinks so that the matrix becomes more regular. 

The regularized estimator for state-dependent autoregressive coefficients is based on the classic LASSO regularization\cite{tibshirani1996regression} such that

$$\V{a}^r=\underset{\V{a}}\arg\min\|\vec(\V{y}_{p+1:T}) - \mu + \sum_{k=1}^p\V{a}_k\Tra\vec(\V{y}_{p+1-k:T-k})\|_2^2 + \lambda_a\|\V{a}\|_1,$$ 
where $\vec$ is the vectorization operator, and $\V{a}=[\V{a}_p\Tra,...,\V{a}_1\Tra]\Tra=[\vec(\M{A}_p)\Tra,...,\vec(\M{A}_1)\Tra]\Tra$ is the vectorization of the state-dependent autoregression matrices. Here $\lambda_a\geq 0$ controls the strength of the regularization on the $\ell_1$ LASSO penalty, i.e. a larger $\lambda_a$ will induce a more sparse estimator.

\subsection{Cross-validation}

The selection of the optimal regularization parameters $\lambda_\Sigma$ and $\lambda_a$ will be based on a similar cross-validation scheme by minimizing one-step-ahead mean-square forecast error (MSFE) as was described in \cite{banbura2010large} and \cite{nicholson2014structured}. More specifically, the data is divided into three periods: one for training (1:$T_1$), one for validation ($T_1$:$T_2$), and one for forecasting ($T_2$:$T$). 

The validation process starts by fitting a model using all data up to time $T_1$ and forecast $\V{y}^{\lambda_\Sigma,\lambda_a}_{T_1+1}$. Then we sequentially add one observation at a time and repeat this process until time $T_2$. Finally, from time T2 to T, we evaluate the one-step-ahead forecast error by minimizing
$$MSFE(\lambda_\Sigma,\lambda_a)=\frac{1}{T_2-T_1}\sum_{t=T_1}^{T_2-1}\|\V{y}^{\lambda_\Sigma,\lambda_a}_{t+1}-\V{y}_{t+1}\|^2_F,$$
where $\|.\|_F$ is the Frobenius norm defined as $\|\M{A}\|_F=\sqrt{\tr(\M{A}\Tra\M{A})}$. A two-dimensional grid-search is adopted to find the regularization values that minimize the MSFE, with 15 grid points in each dimension. 

\subsection{Parameter estimation}

The parameter estimation procedure follows the general framework of EM algorithm for the class of hidden Markov models proposed by \cite{baum1970maximization} and popularized by \cite{dempster1977maximum}. Regarding the implementation of the EM algorithm to maximize the penalized likelihood function, the monotonic property and convergence results have been proved in \cite{green1990use} and \cite{de1995modified}.

In the E-step, the standard forward-backward variables are generalized on the basis of \cite{rabiner1989tutorial} and \cite{yu2010hidden}. Define $$f_{j,n}(\V{y}_{t+1:t+n})=P(\V{y}_{t+1:t+n}|S_{t+1:t+n}=j),$$ i.e. the state-dependent multivariate autoregressive Gaussian density for state j that lasts for duration n.
Then, define the forward variables $$\alpha_t(j,n)=P(S_{t-n+1:t}=j,\V{y}_{1:t} | \V{\theta}),$$ where $j=1,...,M$, $t=1,...,T$, and $n=\{1,...,\min(D,t)\}$. Initialize 
\begin{equation}\alpha_0(j,n)=\delta_j \;\;\;\;\;\; j=1,...,M,
\end{equation} 
Define the recursion
\begin{equation}
\alpha_t(j,n)=\sum_{i=1}^M\sum_{n'=1}^{\min(D,t)}\alpha_{t-n}(i,n') q_{ij}r_i(n) f_{j,n}(\V{y}_{t-n+1:t}).
\end{equation}
Similarly, define the backward variables $\beta_t(j,n)=P(\V{y}_{t+1:T}|S_{t-n+1:t}=j,\V{\theta})$ where $j=1,...,M$, $t=1,...,T$, and $n=\{1,...,\min(D,t)\}$. Initialize $\beta_T(j,n)=1$ and define the recursion
\begin{equation}
\beta_t(j,n)=\sum_{i=1}^M\sum_{n'=1}^{\min(D,T-t)}q_{ji}r_j(n) f_{i,n'}(\V{y}_{t+1:t+n'})\beta_{t+n'}(i,n').
\end{equation}
In addition, define the following 3 sets of auxiliary variables
\begin{equation}
\begin{split} 
\xi_t(i,j)&=P(S_t=i,S_{t+1}=j,\V{y}_{1:T}|\V{\theta}) \\
&=\sum_{n'=1}^{\min(D,t)}\sum_{n=1}^{\min(D,T-t)}\alpha_t(i,n')q_{ij}f_{j,n}(\V{y}_{t+1:t+n})\beta_{t+n}(j,n),
\end{split}
\end{equation}
\begin{equation}
\eta_t(j,n)=P(S_{t-n+1:t}=j,\V{y}_{1:T}|\V{\theta})=\alpha_t(j,n)\beta_t(j,n),
\end{equation}
\begin{equation}
\gamma_t(j)=P(S_t=j,\V{y}_{1:T}|\V{\theta})=\sum_{n=1}^{\min(D,t,T-t)}\eta_t(j,n).
\end{equation}

Then in the E-step, we are ready to compute 
\begin{equation}
\begin{split}
Q(\V{\theta}|\V{\theta}^{(l)})=&E_{\V{\theta}^{(l)}}\left\{\log[P_{\V{\theta}}(\V{y}_1,...,\V{Y}_T,S_1,...,S_T)]|\V{y}_1,...,\V{y}_T\right\}\\
&=E_{\V{\theta}^{(l)}}\left\{\log[P_{\V{\theta}}(S_1,...,S_T)]|\V{y}_1,...,\V{Y}_T\right\}+\\
&E_{\V{\theta}^{(l)}}\left\{\log[P_{\V{\theta}}(\V{y}_1,...,\V{Y}_T|S_1,...,S_T)]|\V{y}_1,...,\V{Y}_T\right\}\\
=&\left[ \sum_{t=1}^T\sum_{i=1}^M\sum_{j\ne i} \frac{\xi_t(i,j)}{\gamma_t(i)} \log q_{ij}\right]+\left[\sum_{i=1}^M \gamma_0(i)\log\delta_i \right]
+\\
&\left[\sum_{t=1}^{T}\sum_{j=1}^{M}\sum_{n=1}^{D}\frac{\eta_t(j,n)}{\gamma_t(i)}\log r_j(n)\right]+\\
&\left[\sum_{t=1}^T\sum_{j=1}^M \gamma_t(j) \log P(\V{y}_t|\V{y}_{t-1:\max(1,t-p)},\V{\mu}_j,\M{\Sigma}_j,\M{a}_j)\right] ,
\end{split}
\end{equation}
where $\V{\theta}^{(l)}$ is the parameter value at the $l^{th}$ iteration, and $P(\V{y}_t|\V{y}_{t-1:\max(1,t-p)},\V{\mu}_j,\M{\Sigma}_j,\M{a}_j)$ is the state-dependent density for the $p^{th}$ order Gaussian autoregressive process.

In the M-step, we can harness the separability of parameters in $Q(\V{\theta}|\V{\theta}^{(l)})$ to maximize each component individually as follows,

\begin{align}
\delta_j &= \gamma_0(j) / \sum_j\gamma_0(j),\\
q_{ij} &=\sum_t\xi_t(i,j) / \sum_{j\ne i}\sum_t\xi_t(i,j),\\
r_j(n)&=\sum_t\eta_t(j,n)/\sum_n\sum_t\eta_t(j,n).
\end{align}

Then $\V{\mu}_j$ is updated as the unpenalized intercept in the weighted least squares regression for the VAR model with LASSO regularization, where each observation $\V{y}_t|\V{y}_{t-1:t-p}$ is weighted by $\gamma_t(j)$. The autoregression matrix $\M{A}_j$ is updated as the coefficients in the same weighted least squares regression with LASSO regularization. These updates are carried out using coordinate descent algorithm detailed by \cite{friedman2007pathwise}.

$\M{\Sigma}_j$ is updated as a convex combination of the weighted error variance from VAR and a scaled identity matrix with the same trace.

\subsection{Asymptotic properties}

The asymptotic properties for the maximum likelihood estimators in HMM under suitable regularity conditions have been proved successively in \cite{leroux1992maximum}, \cite{bickel1998asymptotic}, \cite{douc2001asymptotics}, \cite{cappe2009inference}, and \cite{an2013identifiability}. 

Furthermore, \cite{barbu2009semi} (also in \cite{trevezas2011exact}) extended proof for the consistency and asymptotic normality of the maximum likelihood estimators for finite-state discrete-time hidden semi-Markov models. The conditions and results are summarized as follows,

\begin{enumerate}[topsep=0pt,itemsep=-1ex,partopsep=1ex,parsep=1ex,label=(B\arabic*)]
\item If for any states $i, j \in \{1,...,M\}$, there is a positive integer $\tau$ such that $P(S_{t+\tau}=j | S_t = i) > 0$
\item The conditional state duration distributions $r_i(.)$ have finite support $\forall i \in \{1,...,M\}$.
\end{enumerate}

Under assumptions (B1) and (B2), the maximum likelihood estimator $\hat{\theta}_T$ is strongly consistent as $T \longrightarrow \infty$. 

In the class of hidden semi-Markov model with a finite state space, assumption (B1) means that the Markov chain is irreducible. This holds when all the states communicate with each other, i.e. there is only one communication class in the transition matrix. (B2) automatically holds when we use the discrete nonparametric state duration distribution in the hidden semi-Markov model because we explicit assign probability mass to a finite collection of possible durations. In case a state duration density with infinite support is adopted, we can censor the distribution at a maximum duration D to satisfy the assumption.  

\subsection{Computational cost}

%basically to compute alpha_T, beta_T, then all the computation of other forward backward variables are negligible
To compute the likelihood in the E-step, \cite{rabiner1989tutorial} pointed out that the computational complexity $O(M^2T)$ for an $M$-state HMM with length $T$, and $O(M^2D^2T)$ for an M-state explicit duration HSMM censored at the largest duration D. Further in our VAR(p)-HSMM framework, the dimension of the observed series is $d$ and the order of autoregression is $p$. Therefore, we have to include the computational cost of $O(d^3+d^2p)$ to compute the multivariate normal density in each forward-backward variable. This adds to a total computational cost $O(M^2D^2T(d^3+d^2p))$ in the E-step.

In the $M$-step, the most computationally expensive part is the update for the the autoregression matrices under the elastic net regularization. Based on the results from \cite{friedman2007pathwise}, the computational cost of the coordinate descent algorithm to solve LASSO is $O(Md^2pT)$ for $M$ $p^{th}$ order vector autoregressions of dimension $d$. This computational cost is dominated by that from the E-step.

Therefore, the total computational complexity is $O(M^2D^2T(d^3+d^2p))$ for each EM iteration. As we can see, the algorithm scales linearly in the length of the series $T$ and autoregression order p, but scales quadratically with the number of latent states $M$ and the maximum censored duration $D$, and scales cubically with the dimension $d$.

\section{Analysis on the NYSE portfolio data}

We apply the proposed model on the New York Stock Exchange (NYSE) financial portfolio data, which consists of the daily closing price of 50 most active NYSE stocks from 2015-01-02 to 2016-12-30 so that each time series is of length 504. This data set is publicly available for download in the R package "rarhsmm". We use the log return as the observed multivariate sequence $\{\V{y}_t\}$ with dimension 50 such that

$$y_{t} = \log\frac{\textrm{price}_{t+1}}{\textrm{price}_{t}}\;\;\;\;t=1,...,503,$$

Our analysis shows there is a fairly strong, positive correlation in the lag 0 log returns among most of the 50 stocks. In contrast, the right panel displays the lag 1 correlation matrix, which is rather sparse. Indeed, 83 of the lag 1 sample correlations are significantly different from zero after testing by Fisher z-transformation ($p<0.05$). This sparsity motivates the use of regularized estimators on the state-dependent autoregression matrices in the VAR(p)-HSMM framework. 

The model selection is performed among the competing regularized models [VAR, HMM, VAR(p)-HSMM] using the minimum MSFE criterion. The first 303 observations were used for training, the next 100 for validation, and the final 100 for forecasting. We set 15 grid points that fall with equal space on the log scale between 0.0001 and 1 for LASSO parameter on VAR coefficients. Similarly, we set 15 grid points that fall with equal space on the log scale between 0.1 and 100 for the shrinkage on the covariances. When fitting the VAR-HSMMs, the maximum latent state duration is set to be 30 days and all latent state duration densities are chosen to be discrete nonparametric. 
From Table~\ref{table2}, all competing models perform comparably well in terms of the MSFE. Both regularized VAR(1)-HSMM and VAR(2)-HSMM with 2 states achieved the lowest MSFE of 2.271. Thus, the regularized VAR(1)-HSMM is selected to be the final model since it is more parsimonious. 

\begin{table*}[ht]
\centering
\begin{tabular}{lll }\hline
 Model ID  & Model specification & MSFE \\   \hline 
 1 & Regularized VAR(1) & 2.293 \\
 2 & Regularized HMM with 2 latent states  & 2.288 \\
 3 & Regularized VAR(1)-HSMM with 2 latent states &  2.271  \\
 4 & Regularized VAR(2)-HSMM with 2 latent states &  2.271 \\
 5 & Regularized VAR(1)-HSMM with 3 latent states & 2.289 \\ \hline
 \end{tabular}
\caption{Summary of model selection on the NYSE portfolio data. The regularization parameters are selected using cross-validation by minimizing one-step-ahead mean-square forecast error (MSFE). We set 15 grid points that fall with equal space on the log scale between 0.0001 and 1 for LASSO parameter on VAR coefficients. Similarly, we set 15 grid points that fall with equal space on the log scale between 0.1 and 100 for the shrinkage on the covariances.}
\label{table2}
\end{table*}

\begin{figure}[H]
\includegraphics[width=\columnwidth]{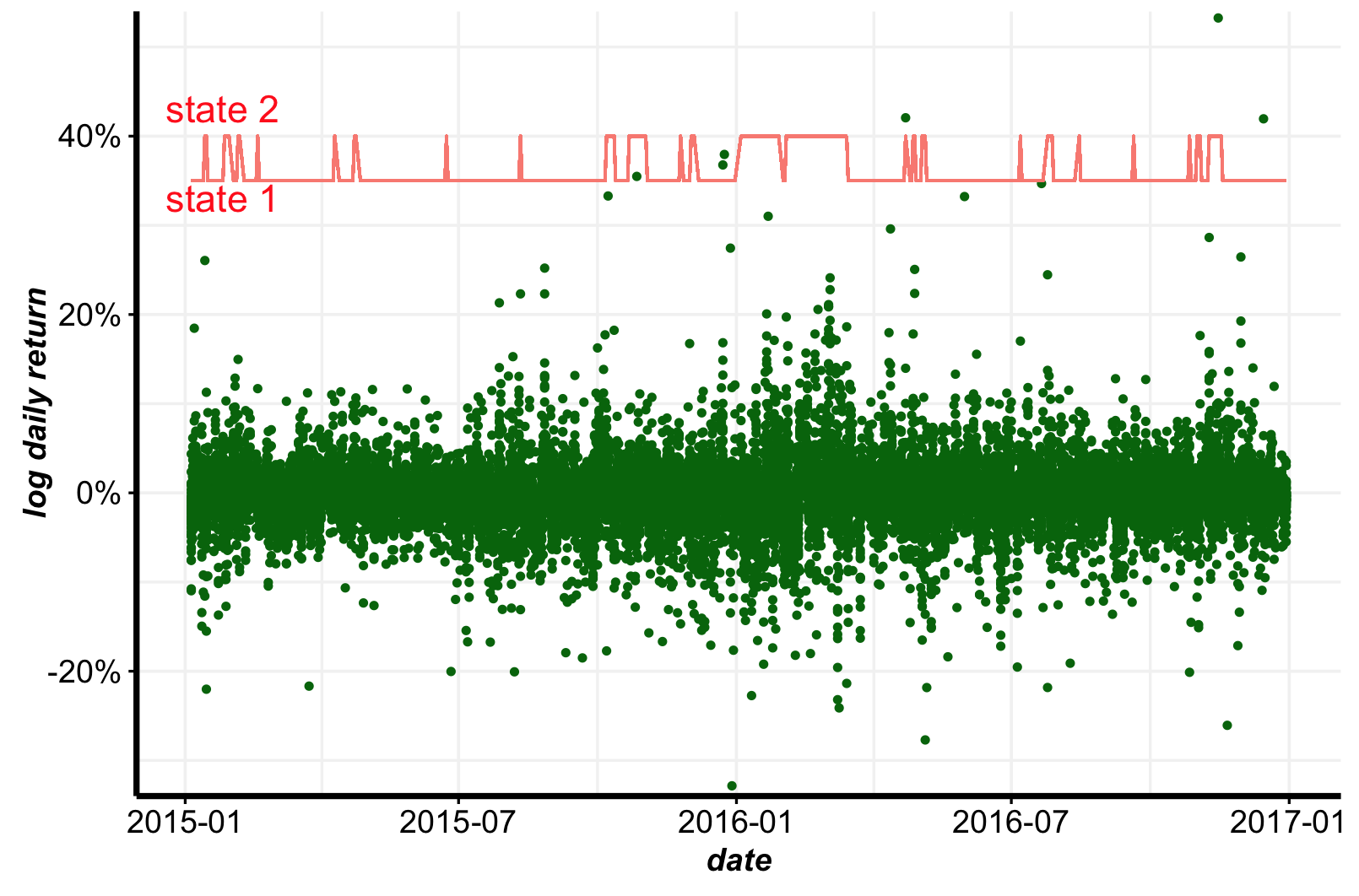}
\caption{Log returns and the decoded latent states for the 50 stocks in the NYSE portfolio from 2015-01-02 to 2016-12-30.}
\label{figure2}
\end{figure}

\begin{figure}[H]
\includegraphics[width=\columnwidth]{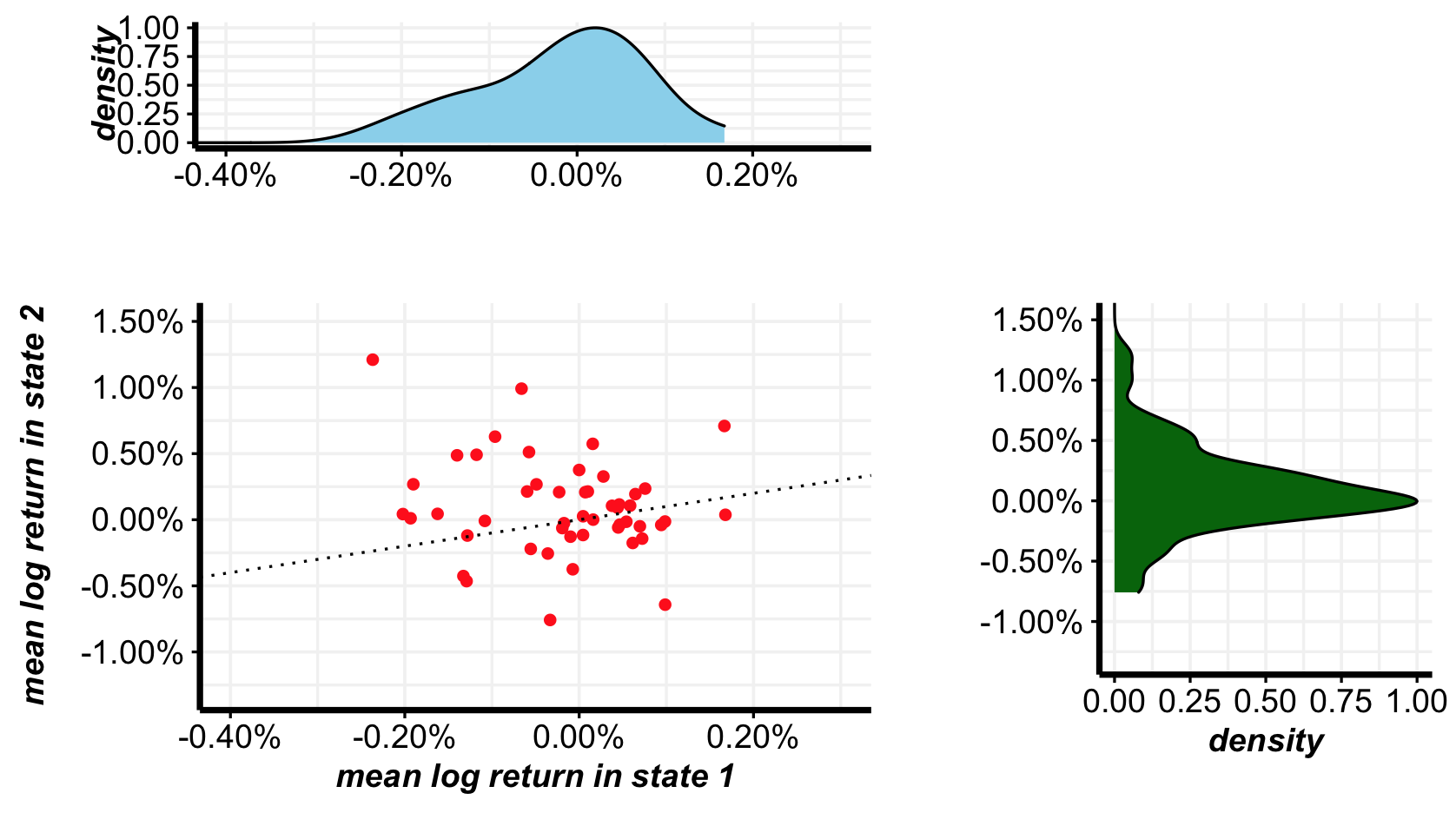}
\caption{Scatter plot and empirical distributions of the fitted means for the log returns in state 1 (stable) and 2 (volatile).}
\label{figure3}
\end{figure}

\begin{figure}[H]
\includegraphics[width=\columnwidth]{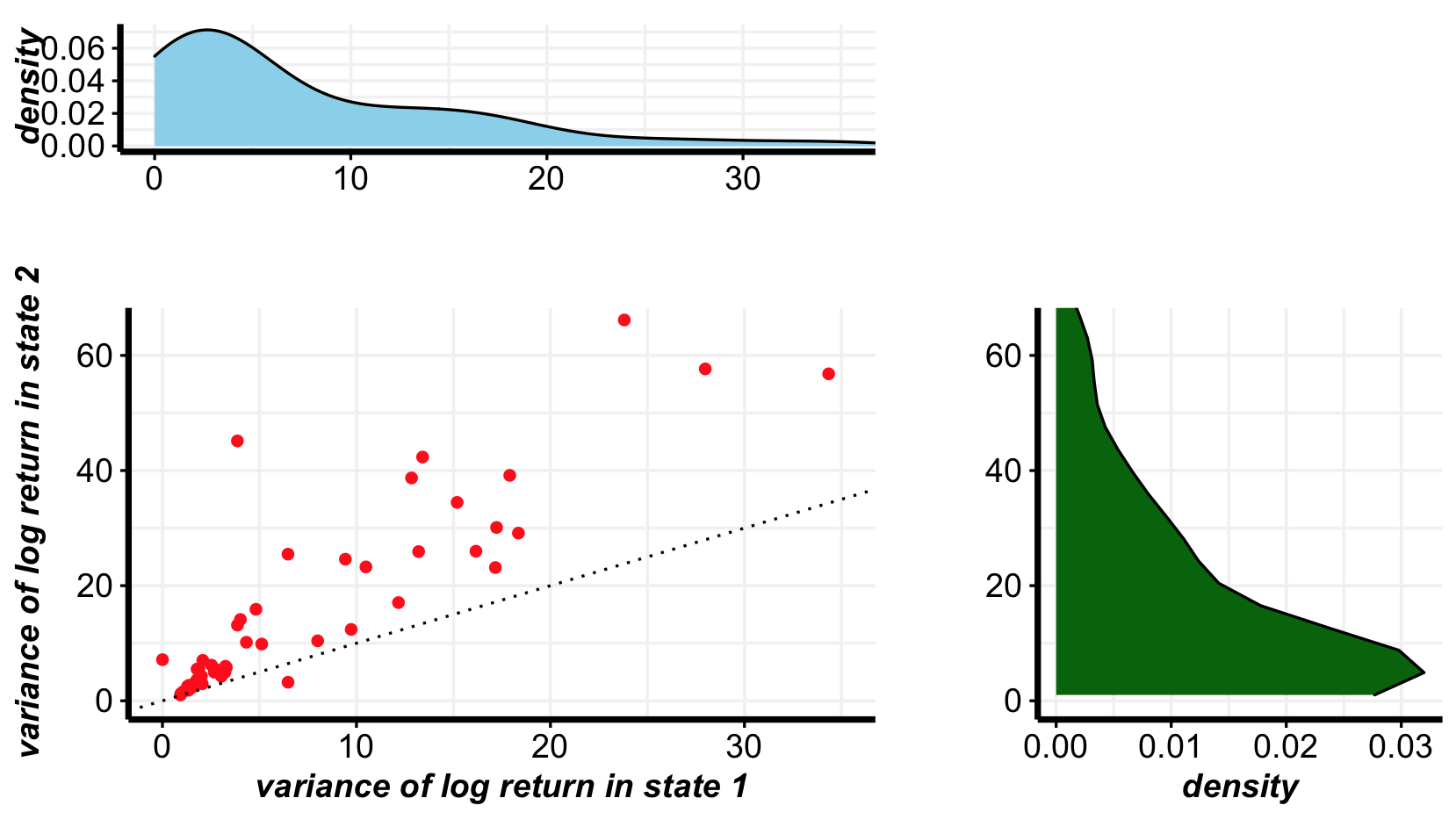}
\caption{Scatter plot and empirical distributions of the fitted variances for the log returns in state 1 (stable) and 2 (volatile).}
\label{figure4}
\end{figure}

The scatter plot in Figure~\ref{figure2} depicts the log returns of the 50 stocks from 2015-01-02 to 2016-12-30. A sequence of the decoded latent states using Viterbi algorithm is overlaid on top of the scatter plot. We can see that state 2 corresponds to the period with a higher volatility in the log return of the 50 stocks while state 1 represents a relatively stable economic period.  
Figure~\ref{figure3} and Figure~\ref{figure4} display the scatter plot and empirical distributions for the fitted means and variances in the two latent states (stable versus volatile). In Figure~\ref{figure3}, we can see that the means in both states are centered around 0, but the spread in the means of state 2 is much larger than that in state 1. In Figure~\ref{figure4}, it seems that most of the stocks have a larger variance for log return in state 2 than in state 1 since the majority of the points lie above the 45 degree line. This result also corroborates the claim that state 2 stands for a more volatile economy than state 1.

%decoded

\section{Discussions}

The class of regularized VAR-HSMM provides a flexible framework to model the switching data generating regimes in multivariate financial time series data, which can work especially well when these state-dependent covariance and autoregression matrices are indeed sparse. In the case study in Section 4, the maximum latent duration (D) is set to be 30 days so as to account for the potential long temporal dependence. We do not want D to be too small, in which case the VAR(p)-HSMM would boil down to VAR(p)-HMM. Although the computation cost of the algorithm increases quadratically in D, the number of parameters only increases linearly in D. In the final regularized model of VAR(1)-HSMM, there are 2909 estimated parameters that are nonzero, where 2550 of them belong to the state-dependent covariance matrices. The fitted means in both states are centered around zero, and there exists strong, positive correlation among most of the stocks in both states. However, the financial returns in state 1 (stable) seems to satisfy the white noise assumption while there is some evidence of lag 1 correlation in state 2 (volatile).

In addition, there are other choices of regularization on the covariance and autoregression matrices. For instance, graphical LASSO \citep[][]{yuan2007model} could be used on the state-dependent covariance matrices and SCAD \citep[][]{fan2001variable} could be used on the autoregression matrices, which is the strategy adopted by \cite{monbet2017sparse} in their VAR-HMM. Another common technique to reduce the number of parameters in covariance and autoregression matrices is to make parametric assumptions on their structures, which will in turn require testing the goodness-of-fit for those assumptions.

%decoded

\bibliographystyle{flairs}

\bibliography{Bibliography-MM-MC}

\newpage

\end{document}